\documentclass{elsart}
\usepackage{graphicx,amssymb}

\newcommand{\be}{\begin{equation}}
\newcommand{\ee}{\end{equation}}
\newcommand{\bea}{\begin{eqnarray}}
\newcommand{\eea}{\end{eqnarray}}
\newcommand{\comment}[1]{}
\newcommand{\fra}[2]{{\textstyle\frac{#1}{#2}}}

\begin{document}
\begin{frontmatter}

\title{Steady states of the conserved Kuramoto-Sivashinsky equation}
\author{Paolo Politi}
\ead{paolo.politi@isc.cnr.it} and
\author{Ruggero Vaia}
\ead{ruggero.vaia@isc.cnr.it}

\address{Istituto dei Sistemi Complessi, Consiglio Nazionale
delle Ricerche, Via Madonna del Piano 10, 50019 Sesto Fiorentino, Italy}

\begin{abstract}
Recent work on the dynamics of a crystal surface [T.~Frisch
and A.~Verga, Phys. Rev. Lett. {\bf 96}, 166104 (2006)] has focused
the attention on the conserved Kuramoto-Sivashinsky (CKS) equation:
$\partial_t u = -\partial_{xx}(u+u_{xx}+u_x^2)$, which displays
coarsening. For a quantitative and qualitative understanding of the
dynamics, the analysis of steady states is particularly relevant. In
this paper we provide a detailed study of the stationary solutions
and their explicit form is given. Periodic configurations form an
increasing branch in the space wavelength--amplitude
($\lambda$--$A$), with $d\lambda/dA>0$. For large wavelength,
$\lambda=4\sqrt{A}$ and the orbits in phase space tend to a
separatrix, which is a parabola. Steady states are found up to
an additive constant $a$, which is set by the dynamics through the
conservation law $\partial_t\langle{u}(x,t)\rangle=0$:
$a(\lambda(t))=\lambda^2(t)/48$.
\end{abstract}

\begin{keyword}
Nonlinear dynamics \sep Coarsening \sep Instabilities \sep Crystal growth
\PACS 05.45.-a \sep 45.20.D- \sep 68.35.Ct \sep 81.10.Aj
\end{keyword}

\end{frontmatter}


\section{Introduction}
\label{intro}

The study of the growth processes of crystal
surfaces~\cite{review,JCG_OPL,Watson} has turned out to be source of
a variety of nonlinear dynamics.
A first, general distinction should be made between a crystal growing
along a high symmetry orientation (e.g., the face (100) of silicon)
and one growing along a vicinal orientation (e.g., the face (119) of
copper). In the former case, growth proceeds~\cite{Evans} via
nucleation, aggregation of diffusing adatoms and coalescence of
islands. The growth is intrinsically two-dimensional and the variable
describing the growth dynamics is the local height $z(x,y,t)$ of the
surface, or alternatively, the local slope $\vec
m(x,y,t)=\nabla{z}(x,y,t)$. The ideal growth of a high-symmetry
surface is called {\it layer-by-layer}. In the latter case (vicinal
orientation), the surface is made up of a train of
steps~\cite{Williams} which capture diffusing adatoms and advance.
Since steps are neither created nor destroyed, a vicinal surface is
usually described in terms of step positions $\zeta_i(x,t)$, where
$i$ labels the steps. In this case the ideal growth is called {\it
step-flow}.

The interest in the nonlinear dynamics of a crystal surface
comes from the observation that growth is often unstable~\cite{libro_JV}.
Layer-by-layer growth may be destabilized by mound formation,
whose phenomenology has many similarities with thermal
faceting~\cite{faceting},
even if the driving force is kinetic rather than thermodynamic.
The representative equation,
\be
\partial_t z = -\nabla\cdot \big[ \nabla^2 \vec m + \vec j (\vec m) \big]~,
\ee
has strict resemblances with the Cahn-Hilliard equation~\cite{Bray},
once we take the gradient of both sides.

The growth of a vicinal surface allows for more rigorous treatments
and the original two-dimensional character of the growth can be often
decomposed in the two spatial degrees of freedom, the coordinate $x$
of the single step and the index $i$ for different steps, according
to the type of emerging instability. In fact, step flow may undergo
two different types of instability: step bunching and step
meandering. Step bunching means that the density of steps does not
keep constant because steps prefer to gather in bunches separated by
large terraces. Step meandering means that steps do not stay straight
and start wandering.

The possibility that both instabilities appear at the same time is
very rare~\cite{Ernst}: typically, step bunching occurs while steps
are straight and step meandering occurs while different steps wander
in phase and their distance remains constant~\cite{JPC}. This means
that the dynamics can be described by effective one-dimensional
equations. For example, step bunching in the presence of large
desorption can be described by the Benney equation~\cite{Uwaha},
\be
\partial_t u = -u_{xx} +\gamma u_{xxx} - u_{xxxx} + u_x^2~,
\ee
giving chaotical dynamics or a regular structure of stable bunches,
depending on the strength of the propagative $\gamma-$term.

The dynamics of meandering depends on the asymmetry in the attachment
kinetics to the steps: meandering occurs when adatoms preferentially
attach to the ascending steps~\cite{BZ}. In the presence of strong
evaporation, we get the well known Kuramoto-Sivashinsky
equation~\cite{BMV}, corresponding to the Benney equation when
$\gamma=0$ and producing spatio-temporal chaos. In the opposite limit
of vanishing evaporation and large asymmetry, a strongly nonlinear
generalization of the Cahn-Hilliard equation was found~\cite{Gillet}:
\be
\partial_t u = -A(u) \partial_{xx} [ B(u) + C(u) u_{xx} ]~.
\ee
This equation may produce~\cite{PRL} a constant-wavelength pattern
($\lambda=\lambda_0$) with diverging amplitude or a coarsening
process, i.e., an increase in time of the the wavelength $\lambda$.
This increase may be either perpetual, therefore defining the
coarsening exponent $n$, $\lambda (t) \sim t^n$, or
interrupted~\cite{Danker} at some length $\lambda_{\mbox{\tiny
max}}$.

The case of vanishing desorption and weak asymmetry has been recently
considered by T.~Frisch and A.~Verga~\cite{FV}, who found the
equation
\be
\partial_t u = -\partial_{xx} ( u+u_{xx}+u_x^2 ) ~,
\label{cks}
\ee
which has been called the {\it conserved Kuramoto-Sivashinsky} (CKS)
equation. The equation has been solved numerically and appears to
give rise to a coarsening pattern, whose exponent, according to
similarity arguments~\cite{FV},
is $n=\fra{1}{2}$. A similar equation, with an
extra propagative term $\gamma u_{xxx}$, arises in step bunching
dynamics with vanishing desorption~\cite{Misbah1} and in a
completely different domain, sand-ripple dynamics~\cite{Misbah2}. The
propagative term breaks the $x\to -x$ symmetry, but does not seem to
change the long-time behavior: perpetual coarsening with
$n=\fra{1}{2}$ is preserved. Additional discussion about this
equation can be found in the last Section.

Most times, coarsening occurs because the Partial Differential
Equation (PDE) describing the
dynamics has a branch of steady states whose wavelength is
an increasing function of their amplitude. These steady states
are unstable with respect to phase fluctuations~\cite{PRE} and the profile
evolves in time keeping close to the stationary branch.

In Ref.~\cite{FV} the authors remark that the profile emerging from
the CKS equation can be thought of as a superposition of parabolas
and that a stationary parabola is a particular solution of the
equation. In this paper we aim at a detailed characterization of the
steady states of the CKS equation. This will be mainly done
analytically, giving the general exact relation between $u$ and
$u_x$, i.e., determining the trajectories in phase space.

\section{Steady states}
\label{sec_ss}

\begin{figure}
\includegraphics*[width=12cm]{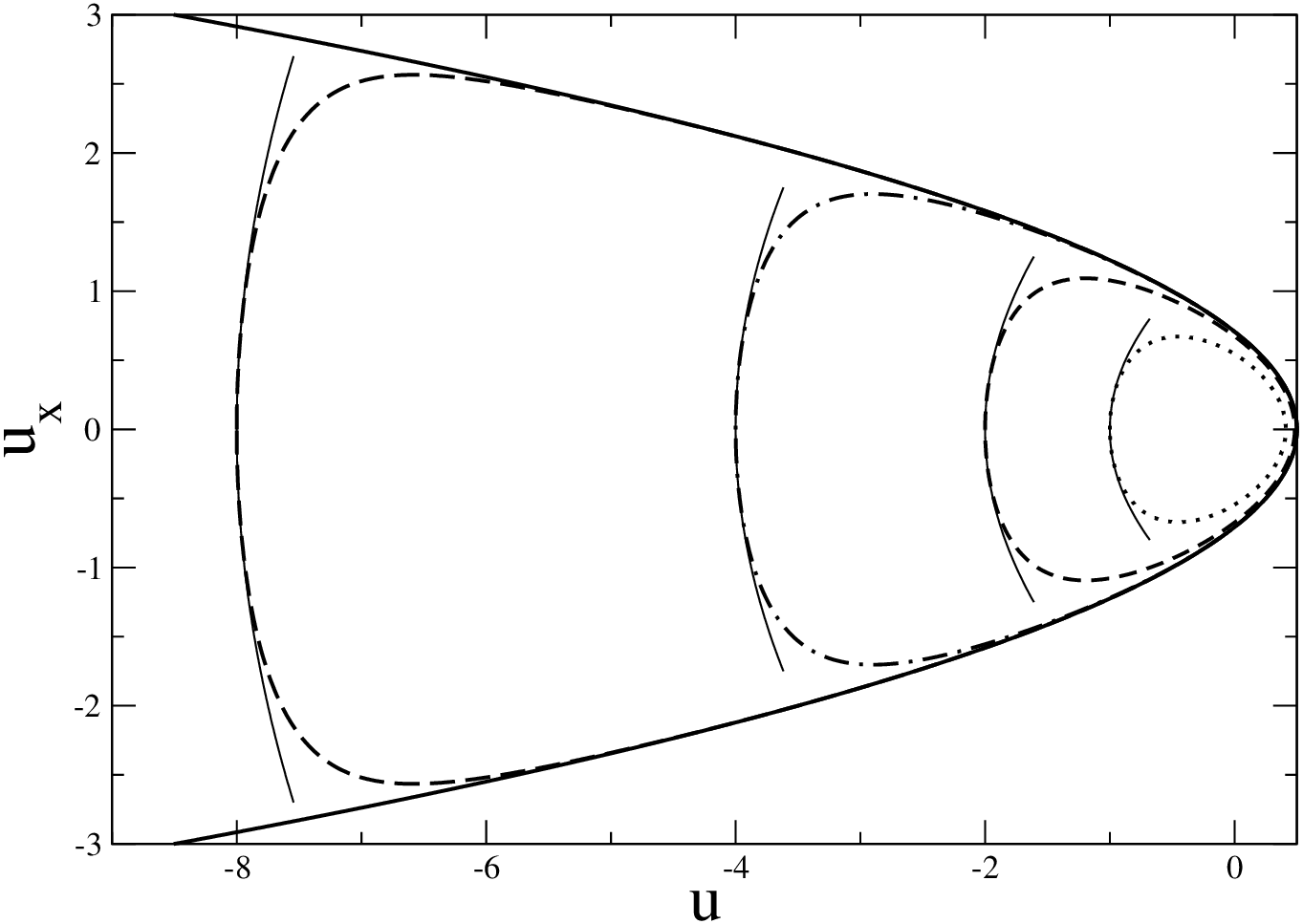}
\caption{Trajectories in the phase space $(u,u_x)$ of the
limiting parabola $u_x^2={1\over 2} - u$ (thick full line) and of the
periodic steady states: $A_-=1$ (dotted), $A_-=2$ (short dashed),
$A_-=4$ (dot dashed), and $A_-=8$ (long dashed). The thin full lines
correspond to the approximation $u_x^2=2A_-(u+A_-)$, valid close to
$u=-A_-$. }
\label{fig_spazio_fasi}
\end{figure}

The steady states of the CKS equation~(\ref{cks}), satisfy the
second-order nonlinear differential equation
\be
 u+u_{xx}+u_x^2 = a + b x ~.
\label{eq_c0}
\ee
Since the constant $b$ must vanish in order to get limited solutions,
while the constant $a$ can be trivially absorbed into a uniform shift
of $u(x)$ and be set to zero, the problem reduces to solving the
differential equation
\be
 u_{xx} = -u - u_x^2 ~.
\label{newton}
\ee
Interpreting $x$ as a time, this equation describes the dynamics of a
harmonic oscillator subject to an additional force proportional to
the square velocity. It can be easily solved by evaluating the
derivative of $u_{xx}\equiv{F}$,
\be
 F_x = -u_x - 2u_x u_{xx} = -u_x (1+2F)~,
\ee
so that
\be
 \frac{du}{dF}= -\,\frac1{1+2F}~,
\ee
whose integration starting from a zero-velocity initial condition,
$u(0)=A$ and $u_x(0)=0$, that implies $F(0)=-A$, gives the first
integral
\be
 2(A-u) = \ln\bigg|\frac{1+2F}{1-2A}\,\bigg| ~,
\ee
or
\be
 1+2F = \pm (1-2A)~ e^{2(A-u)} ~;
\ee
a continuity argument allows us to drop the minus sign
\be
 F = -u-u_x^2 = -\fra12 + (1-2A)~ e^{2(A-u)} ~,
\ee
making it apparent that if $A\ge\fra12$ the force is strictly
negative and the trajectory is not limited. On the other hand, if
$A=A_+$, with $0<A_+<\fra12$\,, the particle initially acquires a
negative velocity, but the increasing positive contribution to the
force $F$ will eventually restore $u_x=0$ at some position
$u=-A_{-}<0$; the converse occurs starting from $A=-A_-<0$, so that a
periodic orbit occurs for any $A<\fra12$. The pairs of turning points
$(A_+,-A_-)$ are solution of
\be
 (1-2A_+)\,e^{-(1-2A_+)}=(1+2A_-)\,e^{-(1+2A_-)}~,
\label{turningpoints}
\ee
that maps the interval $[0,\fra12)$ onto the interval $(-\infty,0]$.
Eventually, we can write the trajectories in the phase space
$(u,u_x)$ as
\be
 u_x^2 = \fra12 -u - \big(\fra12-A\big)\,e^{2(A -u)} ~;
\label{qp}
\ee
these are plotted in Fig.~\ref{fig_spazio_fasi} for a choice of
values of $A_-$. If $A_+=\fra{1}{2}$, we obtain the separatrix
$u_x^2=\fra12-u$, which corresponds to the parabolic trajectory
$u(x)=\fra12-(x-x_0)^2/4$.

\begin{figure}
\includegraphics*[width=12cm]{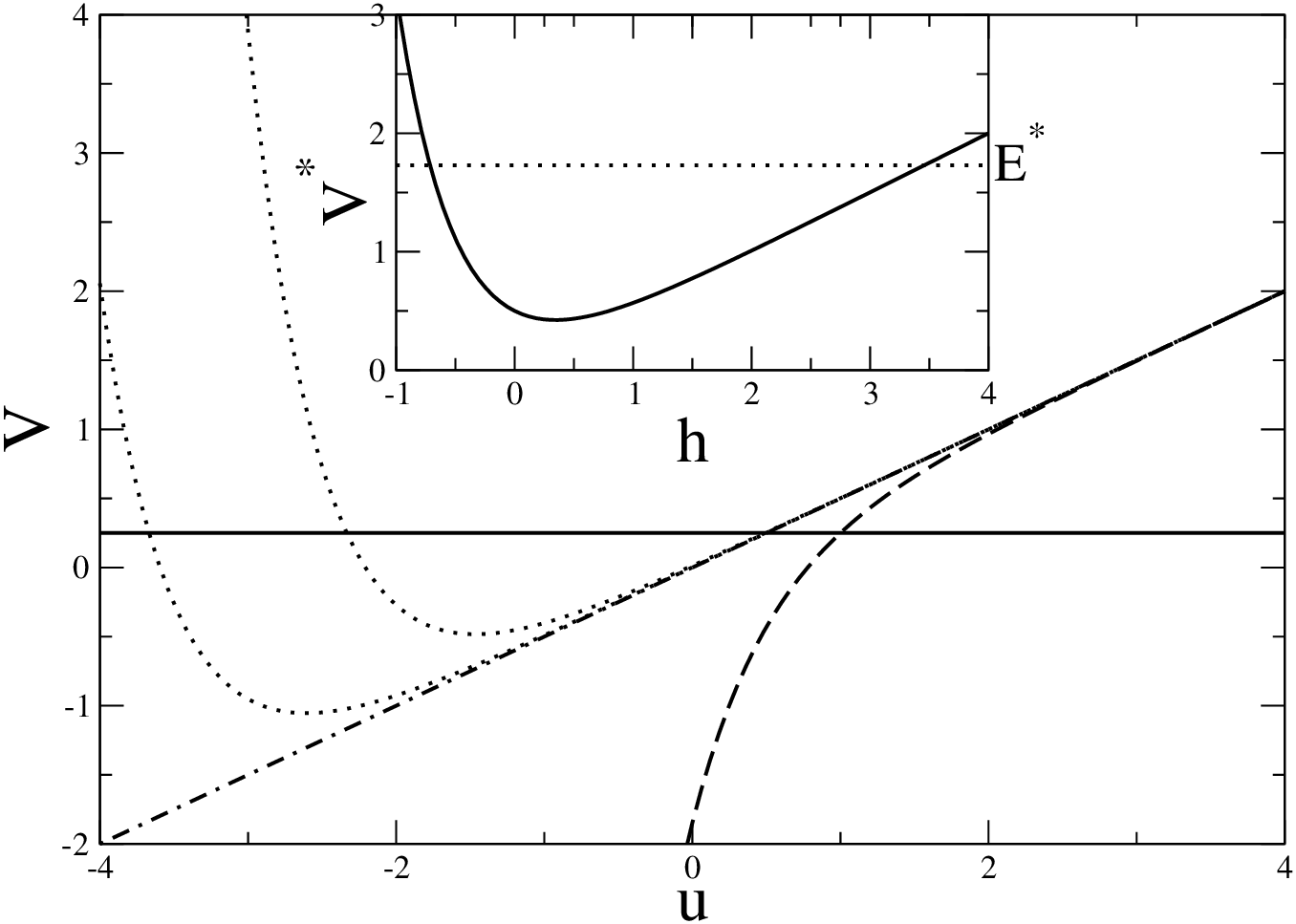}
\caption{Effective potentials $V(u)$ for bound states,
$A<\fra12$ (dotted lines), for the separatrix, $A=\fra12$
(dash-dotted line), and for unbounded states, $A>\fra12$ (dashed
line). The ``energy" of the particle is $E=\fra14$ (full line) in all
cases. The condition $V(u)=E$ identifies the turning points. Inset:
the potential $V^*(h)=\fra12(h+e^{-2h})$ (full line) and the energy
$E^*=\fra12(\fra12-A) -\fra14\ln(\fra12-A)$ (dotted line),
valid for $A<\fra12$. }
\label{fig_potenziale}
\end{figure}

Eq.~(\ref{qp}) can be rewritten in a more familiar form,
\be
 {u_x^2\over 2} + V(u) = E = \fra{1}{4} ~,
\label{first_integral}
\ee
showing that the system satisfies the conservation of a pseudo-energy
$E=\fra{1}{4}$, with a potential
\be
 V(u)= {u\over 2} + \fra{1}{2} (\fra{1}{2}-A)\, e^{2(A-u)}
 \label{Vu}
\ee
that depends on the boundary (initial) conditions. The above
discussion about closed and open orbits becomes a straightforward
analysis of the potential shape, as reported in
Fig.~\ref{fig_potenziale}.

It appears from the figure that the
potential keeps its shape: this can be checked analytically by
considering a shift in the $u$ variable, $u(x)=h(x)+q_0$; for
$q_0=A+\fra12\ln|\fra12-A|$ the first integral (\ref{first_integral})
can be rewritten as
\be
{h_x^2\over 2} + V^*(h) = E^*(A) ,
\ee
with $E^*(A)=\fra12(\fra12-A)-\fra14\ln|\fra12-A|$, and the
$A$-independent potential
\be
 V^*(h) = \fra{1}{2} (h\pm e^{-2h}) ~,
\ee
where the sign $+$ ($-$) is valid for $A<\fra12$ ($A>\fra12$). For
bounded states, the energy minimum is $E^*(0)=\frac14(1+\ln2)$ and
coincides with the minimum of the potential $V^*(h{=}\ln{2}/2)$
(trivial orbit). The potential $V^*(h)$ and a representative value of
the energy $E^*$ are plotted in the inset of
Fig.~\ref{fig_potenziale}.

\begin{figure}
\includegraphics*[width=12cm]{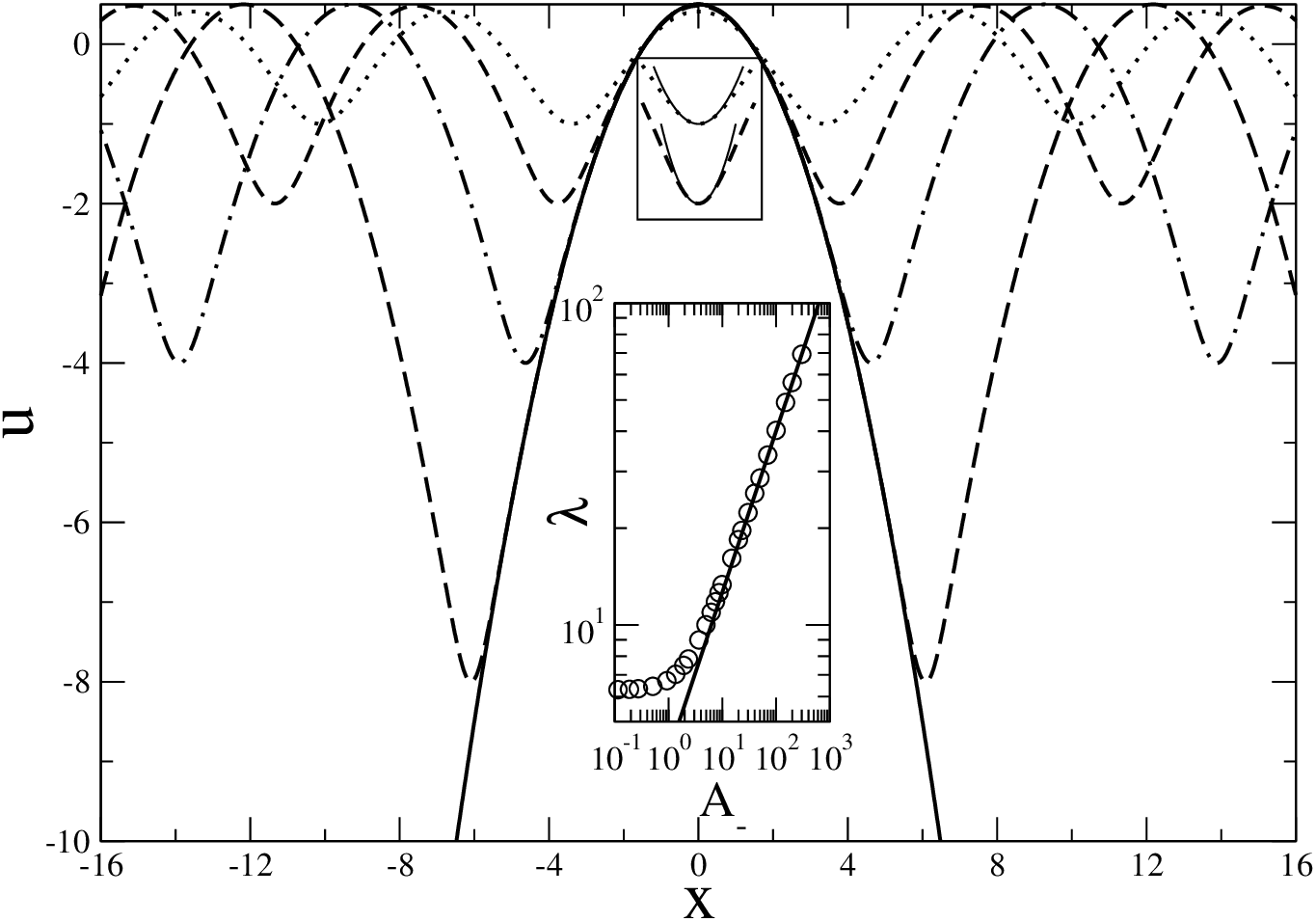}
\caption{Periodic steady states configurations, corresponding to
$A_-=1$ (dotted), $A_-=2$ (short dashed), $A_-=4$ (dot dashed), and
$A_-=8$ (long dashed). The thick full lines corresponds to the
limiting parabola $u={1\over 2} -x^2/4$. Small inset: the minima of
the configurations and the approximations $u=-A_- + A_-x^2/2$ (thin
full lines). Large inset: wavelength $\lambda$ as a function of the
amplitude $A_-$. The full line is the asymptotic relation
$\lambda=4\sqrt{A_-}$. }
\label{fig_traiettorie}
\end{figure}

The numerical results for the periodic configurations $u(x)$ in real
space are shown in Fig.~3. It is useful to determine the amplitude
$A_-$ in the negative $x$ direction, as a function of $A_+$, through
the condition (\ref{turningpoints}). For $A_+\to 0$, it is easily
found that $A_-/A_+\to 1$, while in the important limit
$A_+\to\fra{1}{2}$, $A_-$ diverges logarithmically according to
\be
 A_-e^{-2A_-} \simeq \fra12-A_+ ~,
\ee
i.e.,
\be
 A_- \approx -\fra{1}{2}\ln(1-2A_+) ~.
\ee

In proximity of $u=-A_-$, bounded trajectories have a minimum with a
curvature that diverges as $A_-\to\infty$, as shown by the expansion
$u=-A_-+\varepsilon$ in Eq.~(\ref{qp}), which gives
\be
 u \approx -A_- + {A_-\over 2} (\delta x)^2
 + O\big[A_-^2 (\delta x)^4\big] ~.
\ee
This approximation is shown as thin full lines, both in
Fig.~\ref{fig_spazio_fasi} and in Fig.~\ref{fig_traiettorie}. The
quadratic and quartic terms are of the same order when
$|\delta{x}|\approx{1}/\sqrt{A_-}$, which sets the size of the
high-curvature region. In fact, the slope
$u_x(\delta{x}=1/\sqrt{A_-})\approx\sqrt{A_-}$ joins to the same
slope of the limiting parabola $u=\fra{1}{2}-x^2/4$, when
$u\approx-A_-$.

As for the unbounded states, $A>\frac12$, it is worth noting that
they diverge at a finite value $x_\infty$; in the mechanical analogy,
this means that the particle escapes to infinity in a finite time. We
can determine $x_\infty$ integrating Eq.~(\ref{qp}) once more:
\be
 x_\infty = -\int_{A}^{-\infty} {du\over u_x} =
 {1\over\sqrt{2}}\int_0^\infty {ds\over\sqrt{s+(2A-1)(e^s -1)}}
\ee
and, for large $A$,
\be
 x_\infty \simeq \frac1{\sqrt{4A}}\int_0^\infty \frac{ds}{\sqrt{e^s-1}}
=\frac\pi{\sqrt{4A}}~.
\ee

Finally, in the large inset of Fig.~\ref{fig_traiettorie} we plot the
wavelength $\lambda$ of the steady states as a function of their
amplitude $A_-$. The full line, $\lambda=4\sqrt{A_-}$, gives the
analytical approximation valid for large $A_-$. It can be determined
from the asymptotic parabola $u=\fra{1}{2}-x^2/4$ imposing
$u(\lambda/2)=-A_-$, or equivalently, from the mechanical analogy for
the potential $V^*$ (see inset of Fig.~\ref{fig_potenziale}).

\section{Steady states and dynamics}
\label{sec_dynamics}

In the Introduction we have argued that steady states are important
because dynamics proceeds evolving along the family of steady states
of increasing wavelength $\lambda$. In the case of the CKS equation
special attention should be paid to the constant $a$ appearing in
Eq.~(\ref{eq_c0}) and to the conserved character of Eq.~(\ref{cks}).
We are now going to show that the conservation law fixes the value of
$a$, as a function of $\lambda$. Afterwards, we explain that for the
dynamics $a$ is the time dependent vertical shifting of the surface
profile.

On the one hand, the conserved dynamics requires that the spatial
average $\langle u(x,t) \rangle$ is time independent; on the other
hand, $\langle u(x)\rangle \ne 0$ and does depend on $\lambda$.
Therefore, the family of steady states which is relevant for the
dynamics is
\be
u_\lambda(x) = u(x) + a(\lambda) ~,
\ee
where the constant $a$ satisfies the condition
\be
a(\lambda) = - \langle u(x)\rangle .
\ee

In the above notation, $u(x)$ means the general steady state (of
wavelength $\lambda$) found in the preceding Section. For large
$\lambda$, the average value of $u(x)$ can be safely determined by
approximating it with the arc of the (separatrix) parabola, so that
\be
 \langle u(x) \rangle =
 \frac1\lambda \int_{-\lambda/2}^{\lambda/2} dx\,
 \bigg(\frac12 - {x^2\over 4}\bigg) + o(\lambda^2)
 = -{\lambda^2\over 48} + o(\lambda^2)~.
\ee
Therefore, we get
\be
 a(\lambda) = {\lambda^2\over 48} ~.
\label{a_lambda}
\ee

\section{Conclusions and discussion}

The starting and only assumption of this work is that steady states of
increasing wavelength play a central role in the dynamics of the CKS
equation and, in particular, in the coarsening process.
This assumption is based on previous work~\cite{PRE}
on a plethora of different
PDE for which this connection has been well established.
It is also supported by numerical work~\cite{FV} on the CKS eq., showing that
the system indeed coarsens with a profile made up of a sequence of arcs
of parabola.

In Section~\ref{sec_ss} we have analytically found all the steady
states and in Section~\ref{sec_dynamics} we have determined the
relation between vertical shifting of the profile and coarsening
process. The theory of the phase diffusion equation~\cite{PRE}, which
would allow a more rigorous derivation of the coarsening exponent
$n=\fra12$~\cite{FV}, can not easily be applied to the CKS equation,
because the linear operator coming from the Frech\'et derivative is
not self-adjoint, a feature that the CKS equation shares with the
Kuramoto-Sivashinsky equation.

The fact that the surface profile is made up of arcs of parabola
joined by vanishing regions of diverging amplitude (angular points)
suggests an alternative dynamical description: the system is made up
of a sequence of ``particles" (the angular points) which tend to
annihilate, and therefore the system tends to coarsen. Is it possible
to study the effective dynamics of these particles?

Let us finally comment on a modified version of the CKS
equation~\cite{Misbah1,Misbah2},
\be
\partial_t u = -\partial_{xx} ( u + u_{xx} + u_x^2 +\gamma u_x) ~,
\label{eq_drift}
\ee
where a third-order dispersive drift term has been added to the
right-hand-side. This equation has been mentioned in the Introduction
as emerging in two different domains: growth of a vicinal surface
subject to a step-bunching instability and sand ripple dynamics.
According to numerics~\cite{Misbah1}, Eq.~(\ref{eq_drift}) displays a
surface profile which is not much different from the profiles shown
by the CKS equation. Also, the system seems to coarsen with the same
law, $\lambda \sim \sqrt{t}$.

It may be tempting to treat Eq.~(\ref{eq_drift}) along the same lines
followed for the CKS equation. The first remark is that periodic
steady states should now be replaced by travelling periodic
configurations. In fact, Eq.~(\ref{eq_drift}) does not have periodic
stationary solutions because of the $\gamma-$term. Nonetheless, one
might look for periodic solutions depending on $(x-vt)$. It would be
interesting to understand if there is a branch of travelling periodic
configurations in the space $\lambda-A$ and if dynamics coarsens
following this branch, as for the CKS equation.

\ack
We acknowledge useful discussions with T. Frisch, S. Lepri and A. Verga.

\end{document}